\documentclass[journal,twoside,web]{IEEEtran}

\usepackage[ruled,algoruled,noend,noline]{algorithm2e}
\SetKwIF{If}{ElseIf}{Else}{if}{}{else if}{else}{end if}%
\usepackage{amsfonts,amsmath,amssymb}
\usepackage{balance}
\usepackage{bm}
\usepackage{caption}
\usepackage{cite}
\usepackage{enumitem}
\usepackage[T1]{fontenc}
\usepackage{gensymb}
\usepackage{graphicx}
\usepackage[utf8]{inputenc}
\usepackage{longtable}
\usepackage{mathrsfs}
\usepackage{mathtools}
\usepackage{multirow}
\usepackage[cmintegrals]{newtxmath}
\usepackage{siunitx}
\usepackage{soul}
\usepackage{subfigure}
\usepackage{tabularx}
\usepackage{textcomp}
\usepackage[color=green!40]{todonotes}
\soulregister\cite7
\soulregister\footnote7
\soulregister\eqref7
\soulregister\ref7
\usepackage{threeparttable}
\usepackage{url}
\usepackage{wrapfig}
\usepackage{xcolor}
\usepackage[colorlinks,bookmarksopen,bookmarksnumbered,citecolor=red,urlcolor=red]{hyperref}
\usepackage{hyperref}
\hypersetup{hidelinks}
\newcommand{\bigO}{\mathcal{O}}
\newcommand{\vect}[1]{\boldsymbol{\mathbf{#1}}} % see:
\newcommand{\slfrac}[2]{\left.#1\middle/#2\right.}
\DeclareMathOperator*{\diag}{diag}

\graphicspath{{./}{figures/}{../figures/}}

\newcommand{\orcidA}[1]{\href{https://orcid.org/0000-0002-8382-785X}{\includegraphics[width=10pt]{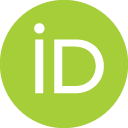}}}%
\newcommand{\orcidB}[1]{\href{https://orcid.org/0000-0003-4231-7455}{\includegraphics[width=10pt]{ORCIDiD_icon.png}}}%
\newcommand{\orcidC}[1]{\href{https://orcid.org/0000-0002-4123-2647}{\includegraphics[width=10pt]{ORCIDiD_icon.png}}}%
\newcommand{\orcidD}[1]{\href{https://orcid.org/0000-0002-0212-2365}{\includegraphics[width=10pt]{ORCIDiD_icon.png}}}%
\newcommand{\orcidE}[1]{\href{https://orcid.org/0000-0003-4175-7248}{\includegraphics[width=10pt]{ORCIDiD_icon.png}}}%
\newcommand{\myauthor}{%
  Zhe Tang\orcidA{0000-0002-8382-785X},~\IEEEmembership{Member,~IEEE},~%
  Sihao Li\orcidB{0000-0003-4231-7455},~%
  Zichen Huang\orcidE{0000-0003-4175-7248},~%
  Guandong Yang,~%
  Kyeong Soo
  Kim\orcidC{0000-0002-4123-2647},~\IEEEmembership{Senior~Member,~IEEE},~%
  Jeremy S. Smith\orcidD{0000-0002-0212-2365},~\IEEEmembership{Member,~IEEE},~%
  Zhaowei Zhu\orcidD{0000-0003-3894-5862},~\IEEEmembership{Member,~IEEE},~%
  and Qi Xuan\orcidD{0000-0002-1042-470X},~\IEEEmembership{Senior~Member,~IEEE}
}%
\newcommand{\mytitle}{Decentralized Indoor Localization Based on A Sparse
  Gaussian Process with Reduced-Dimensional Inputs for Real-Time Sensing and
  Training on IoT Devices}%

\begin{document}

\title{\mytitle}

\author{%
  \myauthor%
  \thanks{%
    Z. Tang and Q. Xuan are with the Institute of Cyberspace Security, College
    of Information Engineering, Zhejiang University of Technology, Hangzhou
    310000, China, and also with the Binjiang Institute of Artificial
    Intelligence, Zhejiang University of Technology, Hangzhou 310000, China
    (e-mail: tangzhe@bcszjut.com; xuanqi@zjut.edu.cn).}%
  \thanks{%
    S.~Li is with the School of Artificial Intelligence, Suzhou Institute of
    Industrial Technology, Suzhou 215123, China (e-mail: 01177@siit.edu.cn).}%
  \thanks{%
    Z.~Huang, G.~Yang, and K.~S.~Kim are with the School of Advanced Technology,
    Xi'an Jiaotong-Liverpool University, Suzhou 215123, China (e-mail:
    Zichen.Huang20@alumni.xjtlu.edu.cn; Guandong.Yang21@alumni.xjtlu.edu.cn;
    Kyeongsoo.Kim@xjtlu.edu.cn).}%
  \thanks{%
    J.~S.~Smith is with the Department of Electrical Engineering and
    Electronics, University of Liverpool, Liverpool L69 3GJ, U.K. (e-mail:
    J.S.Smith@liverpool.ac.uk).}%
  \thanks{%
    Z.~Zhu is with the Binjiang Institute of Artificial Intelligence, Zhejiang
    University of Technology, Hangzhou 310000, China, and also with the D5 Data
    Co., Ltd., Hangzhou 310000, China (e-mail: zzw@d5data.ai).}%
}%%

\maketitle

\begin{abstract}
  As a large number of Internet of Things (IoT) devices are deployed in the
  field, there arises huge potential of edge computing for indoor localization
  on those devices. Conventional indoor localization based on a centralized
  server with substantial computational resources, often covering a number of
  multistory buildings, cannot easily adapt to time-varying indoor
  electromagnetic environments due to its high cost of fingerprint database
  update and model retraining; the centralized server is also susceptible to
  security breaches. To address these issues, we propose a decentralized indoor
  localization framework, leveraging models based on a Sparse Gaussian Process
  with Reduced-dimensional Inputs (SGP-RI) deployed to IoT devices for a smaller
  service area, which can quickly adapt to time-varying indoor electromagnetic
  environments through real-time sensing and retraining. The experimental
  results based on a multibuilding, multifloor static database and a
  single-building, single-floor dynamic database, demonstrate the feasibility of
  the proposed framework, where the SGP-RI with less than half the training
  samples can produce localization performance comparable to the standard
  Gaussian process (GP) with the whole training samples.
\end{abstract}


\IEEEpeerreviewmaketitle

\section{Introduction}
\label{introduction}
%%%
\IEEEPARstart{W}{i-Fi} fingerprinting has emerged as a pivotal technique for
indoor localization, leveraging the existing Wi-Fi infrastructure without the
necessity of supplementary hardware. Conventional Wi-Fi fingerprinting consists
of \textit{offline phase}, where fingerprint databases are constructed and a
model is trained based on them, and \textit{online phase}, where the location of
a user or a device is estimated using the trained model with newly-measured data
at an unknown location~\cite{s21238086}. For the construction of fingerprint
databases, Received Signal Strengths (RSSs), Received Signal Strength Indicators
(RSSIs), or Channel State Information are collected at Reference Points (RPs)
covering service areas, which is labor-intensive and time-consuming, spurring an
investigation to alleviate the burdens of fingerprint database
construction~\cite{Low-Cost-Method-A-Review}.

After the fingerprint database construction, a localization algorithm is
selected based on the application's context. Of many candidates, deep learning
has been extensively used, which is based on Deep Neural Network
(DNN)~\cite{DNN}, Convolutional Neural Network (CNN)~\cite{CNNLoc}, Recurrent
Neural Network (RNN)~\cite{2021hierarchical},
Transformer~\cite{Transformers-indoor-localization}, Graph Neural Network
(GNN)~\cite{10601173}, and their hybrid~\cite{CNN-Transformer}. The training of
these models requires powerful servers equipped with Graphics Processing Units
(GPUs) and large storage space. Though researchers try to reduce the complexity
and training cost of a model (e.g., reducing the feature space
dimension~\cite{10601173}), most indoor localization systems rely on centralized
models running on powerful
servers~\cite{DNN,CNNLoc,2021hierarchical,Transformers-indoor-localization,10601173,CNN-Transformer}.

Note that a significant amount of computational resources is available on a
large number of network and Internet of Things (IoT) devices deployed in the
field. For instance, Wireless Access Points (WAPs) running
OpenWrt~\cite{OpenWrt} and its many variations can provide most of the
functionalities available on general server platforms in addition to
network-specific ones. Likewise, IoT devices such as Raspberry Pi, which are
equipped with powerful processors, USB ports, and network interfaces, can
provide computational resources comparable to those of desktop computers and
servers~\cite{venkatesh2018iot}. Although distributed and limited in
computational power and storage space compared to those of centralized servers,
the redundant computational resources on those network and IoT devices could be
exploited for indoor localization.

It is in this context that we propose a new decentralized indoor localization
framework leveraging edge computing, under which real-time-trainable IoT indoor
localization models based on a Sparse Gaussian Process with Reduced-dimensional
Inputs (SGP-RI) are deployed to resource-constrained IoT devices. Decentralized
indoor localization has the major advantages in terms of reliability and privacy
over centralized one: As for reliability, the failure of a model running on one
device affects only the service area it covers unlike a centralized server that
is a \textit{single point of failure}. Regarding privacy, a centralized server
handles location requests from users across all buildings and floors, which
could breach the privacy of the users by tracking their locations and/or
trajectories; under decentralized indoor localization, the privacy issue could
be limited to a much smaller service area covered by each localization model.
Compared to the prior works on decentralized indoor localization, which focus on
optimizing algorithms for edge devices (e.g.,
\cite{8454750,kim25:_dual_prong_solut_accur_decen_tag_system}), handling device
heterogeneity (e.g., \cite{ye22:_edgel}), and exploiting distributed
localization (e.g., \cite{nikitaki12:_decen}), the unique contribution of our
proposal is the adaptability of its comprehensive framework and thereby
providing a solution targeting dynamic environments in the context of Wi-Fi
fingerprinting. Specifically, the proposed decentralized indoor localization can
provide on-demand indoor localization services even in large-scale multibuilding
and multifloor environments and quickly adapt to time-varying indoor
electromagnetic environments through real-time sensing and training of the
SGP-RI models deployed on edge devices, each of which independently covers a
much smaller service area.

The rest of the paper is organized as follows: Section~\ref{sec:challenges}
discusses the challenges in training models on resource-constrained IoT
devices. Section~\ref{sec:localization-model} describes the proposed SGP-RI
model and the decentralized indoor localization
framework. Section~\ref{sec:exp-results} presents the results of the performance
evaluation of the proposed SGP-RI model in comparison to reference models on
both a centralized server and an IoT device. Section~\ref{sec:conclusions}
concludes our work.

\section{Challenges in IoT-Based Indoor Localization}
\label{sec:challenges}
%%%
A variety of algorithms and techniques have been proposed for indoor
localization based on resource-constrained platforms: In ~\cite{garg2023sirius},
triangulation using a special implementation of the Angle of Arrival (AoA)
without synchronization was developed for IoT sensors, while efficient solution
techniques of trilateration were studied for resource-constrained devices
in~\cite{8454750}. The pre-training methods and machine learning frameworks
proposed for resource-constrained platforms in~\cite{8953024} and~\cite{9149180}
can be applied to IoT-based indoor localization.

In Wi-Fi fingerprinting, not only the efficacy of algorithms and models but also
the quantity and quality of fingerprint data are critical for localization
performance. Data-sparse approaches like Bayesian models are capable of being
trained with fewer training data for reasonable performance~\cite{8054360} than
data-intensive approaches, such as Neural Networks (NNs). Given the limited
computational resources and storage capacity of IoT devices, striking the right
balance between localization performance and resource utilization, including the
amount of data, is of vital importance for IoT-based indoor localization, which
makes data-sparse approaches more attractive.

As for fingerprint databases, the large numbers of RPs and WAPs in a database do
not necessarily guarantee the quality of the database.
In~\cite{moghtadaiee2014design}, it was demonstrated that the spatial
distribution of RPs, governed by their inter-spacing and alignment, has a more
profound impact on the localization performance than their number. The results
suggest that pre-processing of WAPs and/or RPs is essential to IoT-based indoor
localization, including the handling of undetected WAPs in Wi-Fi RSSI
fingerprint databases~\cite{li24:_wi_fi}, whose RSSI values are set to an
extrapolated minimum for their continuation with the rest of the data (e.g.,
-110 for the UJIIndoorLoc database~\cite{UJI,DNN}).

Those extrapolated values in databases, however, may worsen model
performance. Gaussian Process (GP), for example, imposes strict requirements on
the quality and distribution of data; extrapolated data may skew its
kernel. Such distortion may bias the model's interpretation and judgment of the
data, potentially resulting in the failure to fit the correct hyperplane. Given
the cubic computational complexity of GP, minimizing the number of extrapolated
values is particularly crucial when the dataset size is inherently
constrained~\cite{GPML}. In general, reducing the number of extrapolated values
for non-parametric models is vital to preserving model performance and
efficiency~\cite{tran2016variational}, which is especially the case for
IoT-based indoor localization.

\section{Localization Model Based on SGP-RI}
\label{sec:localization-model}
%%%
To enable real-time training on resource-constrained IoT devices, we base our
indoor localization model on GP regression formulated as a Bayesian linear
model, which does not require backpropagation used to train highly-nonlinear
NNs. Even without the backpropagation, GP regression still suffers from its
cubic scalability with the size of the training dataset. We tackle the issue of
poor scalability by approximating GP using SGP-RI.

\subsection{From GP to SGP}
\label{sec:gp-to-sgp-rdip}
%%%
\subsubsection{GP}
\label{sec:GP}
%%%
A GP is a stochastic process defined over a continuous domain as
follows~\cite{GPML}:
\begin{equation}
  \label{eq:gp}
  f(\vect{x}) \sim \mathcal{GP}(m(\vect{x}), k(\vect{x}, \vect{x}')),
\end{equation}
where $m(\vect{x})$ is the mean function and $k(\vect{x},\vect{x}')$ is the
covariance, also called kernel function, evaluated at $\vect{x}$ and
$\vect{x}'$.

Consider a training dataset $\mathcal{D}{=}\{(\vect{x}_i,y_i)\}_{i=1}^{N}$,
where $\vect{x}_{i}$ is a $W$-dimensional column input vector and $y_{i}$ is a
scalar output. As we cannot directly observe a function value $f(\vect{x})$, its
observation $y$ can be modeled as follows:
\begin{equation}
  \label{eq:gp-noise-model}
  y = f(\vect{x}) + \epsilon,
\end{equation}
where $\epsilon$ is additive white Gaussian noise with variance $\sigma^{2}$,
i.e., $\epsilon{\sim}\mathcal{N}(0,\sigma^{2})$. In Wi-Fi RSSI fingerprinting,
$\vect{x}_{i}$ is a vector of RSSIs from $W$ WAPs measured at the $i$th RP, and
$y_{i}$ is one of the coordinates of the $i$th RP, which means that we need two
GPs for two-dimensional (2D), single-floor localization.\footnote{2D
  localization can be formulated using a single Multioutput Gaussian Process
  (MOGP)~\cite{MOGP-sensors}, but at the expense of increased complexity of
  kernel formation.} The feature part of the training dataset $\mathcal{D}$ can
be represented by an $N{\times}W$ matrix
$\vect{X}{=}[\vect{x}_1,\vect{x}_2,{\ldots},\vect{x}_N]^{\intercal}$, where each
row corresponds to the feature vector of a data point. Likewise, the outputs are
collected in an $N$-dimensional column vector
$\vect{y}{=}[y_{1},y_{2},{\ldots},y_{N}]^{\intercal}$.

The predictive distribution of the test outputs $\vect{f}_{*}$ at test inputs
$\vect{X}_{*}$ is given by
%%%
\begin{equation}
  \label{eq:gp-posterior}
  p(\vect{f}_* | \vect{X}_*, \mathcal{D}) = \mathcal{N}(\vect{\mu}_*, \vect{\Sigma}^2_*)
\end{equation}
%%%
with
%%%
\begin{align}
  \label{eq:gp-posterior-mean}
  \vect{\mu}_* & = \vect{K}_{\vect{X}_*\vect{X}} [\vect{K}_{\vect{XX}} +
                 \sigma^2\vect{I}]^{-1} \vect{y} , \\
  \label{eq:gp-posterior-cov}
  \vect{\Sigma}^2_* & = \vect{K}_{\vect{X}_*\vect{X}_*} - \vect{K}_{\vect{X}_*\vect{X}} [\vect{K}_{\vect{XX}} + \sigma^2\vect{I}]^{-1}\vect{K}_{\vect{X}\vect{X}_{*}} ,
\end{align}
%%%
where $\vect{K}_{\vect{XX}}$, $\vect{K}_{\vect{X}\vect{X}_{*}}$,
$\vect{K}_{\vect{X}_{*}\vect{X}}$, and $\vect{K}_{\vect{X}_*\vect{X}_*}$ are the
covariance matrix for the training inputs, the covariance matrix between the
training and the test inputs, the covariance matrix between the test and the
training inputs, and the covariance matrix for the test inputs, respectively,
and $\vect{I}$ is the identity
matrix. \eqref{eq:gp-posterior}--\eqref{eq:gp-posterior-cov} constitute the
canonical form of GP regression, and their time complexity is $\bigO(N^3)$ with
respect to the number of training data $N$, which is dominated by the inversion
of $[\vect{K}_{\vect{XX}}{+}\sigma^2\vect{I}]$ through Gaussian elimination or
Cholesky decomposition~\cite{GPML}.

\subsubsection{SGP}
\label{sec:SGP}
%%%
The time complexity of $\bigO(N^3)$ makes the application of GP to problems with
large datasets impractical. To address the scalability issue, various techniques
have been proposed to approximate GP based on SGP, which incorporates a small
number $M$ ($M{\ll}N$) of inducing points to reduce the computational complexity
of the GP. The primary distinction among SGP models, therefore, lies in the
selection of inducing points
$\vect{Z}{=}[\vect{z}_{1},\vect{z}_{2},{\ldots},\vect{z}_{M}]^{\intercal}$~\cite{quinonero2005unifying}.

Once inducing points are selected, we can evaluate the values of $f$ at those
inducing points, which are called inducing variables, i.e.,
$\vect{u}{=}[u_{1},u_{2},{\ldots},u_{M}]^{\intercal}$, where
$u_{i}{=}f(\vect{z}_{i})$. Variational inference is then employed to construct
the approximate posterior distribution based on inducing points and inducing
variables, which, in turn, allows us to derive the approximate posterior
distribution of $\vect{f}_*$ at $\vect{X}_*$.

The posterior distribution of $\vect{u}$ given $\mathcal{D}$ and $\vect{Z}$ is
as follows~\cite{SGP_pseudoinput}:
%%%
\begin{equation}
  \label{eq:SGP_instead}
  \begin{multlined}
    p(\vect{u}|\mathcal{D},\vect{Z}) = \\
    \mathcal{N}(\vect{u}|\vect{K}_{\vect{ZZ}}\vect{Q}_{\vect{ZZ}}^{-1}\vect{K}_{\vect{ZX}}[\vect{\Lambda}+\sigma^2\vect{I}]^{-1}\vect{y},
    \vect{K}_{\vect{ZZ}}\vect{Q}_{\vect{ZZ}}^{-1}\vect{K}_{\vect{ZZ}}),
  \end{multlined}
\end{equation}
%%%
with
%%%
\begin{align}
  \label{eq:Q_ZZ}
  \vect{Q}_{\vect{ZZ}} & = \vect{K}_{\vect{ZZ}} + \vect{K}_{\vect{ZX}}[\vect{\Lambda} +
                         \sigma^2\vect{I}]^{-1}\vect{K}_{\vect{XZ}}, \\
  \label{eq:Lambda}
  \vect{\Lambda} & = \diag(\lambda_{1},\ldots,\lambda_{N}),
\end{align}
%%%
where
%%%
\begin{equation}
  \label{eq:lambda}
  \lambda_{i} = [\vect{K}_{\vect{XX}} - \vect{K}_{\vect{XX}}^{\intercal}\vect{K}_{\vect{ZZ}}^{-1}\vect{K}_{\vect{XX}}]_{i,i}.
\end{equation}
%%%
In this case, the predictive distribution of the test outputs $\vect{f}_{*}$ at
test inputs $\vect{X}_{*}$ given $\mathcal{D}$ and $\vect{Z}$, is given by
%%%
\begin{equation}
  p(\vect{f}_{*}|\vect{X}_{*},\mathcal{D},\vect{Z}) = \mathcal{N}(\vect{f}_{*}|\vect{\tilde{\mu}}_{*},\vect{\tilde{\Sigma}}^{2}_{*}),
  \label{eq:SGP-pred}
\end{equation}
%%% 
where
%%% 
\begin{align}
  \vect{\tilde{\mu}}_{*} & =
                           \vect{K}_{\vect{X}_*\vect{X}_*}^{\intercal}\vect{Q}_{\vect{ZZ}}^{-1}\vect{K}_{\vect{ZX}}(\vect{\Lambda}
                           + \sigma^2\vect{I})\vect{y}, \\
  \vect{\tilde{\Sigma}}^{2}_{*} & =\vect{K}_{\vect{X}_*\vect{X}_*}-\vect{K}_{\vect{X}_*\vect{X}_*}^{\intercal}(\vect{K}_{\vect{ZZ}}^{-1}-\vect{Q}_{\vect{ZZ}}^{-1})\vect{K}_{\vect{X}_*\vect{X}_*}.
\end{align}
%%%
The means (i.e., $\vect{\tilde{\mu}}_*$) and the variances (i.e., the diagonal
elements of $\vect{\tilde{\Sigma}}^2_*$) of the distribution provide the
estimates of the test outputs $\vect{f}_*$ and their uncertainties,
respectively.

The time complexity of SGP is significantly reduced to $\bigO(NM^2)$ compared to
$\bigO(N^{3})$ of GP given $M{\ll}N$ because we can avoid the inversion of the
$N{\times}N$ matrix in \eqref{eq:gp-posterior-mean} and
\eqref{eq:gp-posterior-cov}; the major component is the calculation of
$\vect{Q}_{\vect{ZZ}}$ in \eqref{eq:Q_ZZ} dominated by the matrix multiplication
of
$\vect{K}_{\vect{ZX}}[\vect{\Lambda}{+}\sigma^2\vect{I}]^{-1}\vect{K}_{\vect{XZ}}$,
where the complexity of the inversion of the diagonal matrix
$[\vect{\Lambda}{+}\sigma^2\vect{I}]$ is negligible~\cite{SGP_pseudoinput}.

\subsection{SGP with Reduced-Dimensional Inputs (SGP-RI)}
\label{sec:sgp-ri}
%%% 

\subsubsection{WAP-Based Feature Selection}
\label{sec:wap-filtering}
%%%
With SGP based on $M$ inducing variables, we can reduce the computational
complexity of the regression to $\bigO(NM^{2})$. To enable real-time training of
and inference with the SGP regression model on resource-constrained IoT devices,
we go one step further to reduce the dimensionality of input data (i.e., $W$),
which is hardly taken into account in the analysis of computational complexity
of GP models in the literature. For example, the complexity of the calculation
of the covariance matrix $\vect{K}_{\vect{ZZ}}$ in \eqref{eq:SGP_instead} is
$\bigO(M^{2}W)$ because there are $M{\times}M$ elements, each of which depends
on two $W$-dimensional vectors. Though the various calculations depending on $W$
are not captured in the computational complexity of SGP regression, they still
affect the execution time of a model running on an IoT device.

For dimensionality reduction in Wi-Fi fingerprinting, we propose a simple,
heuristic feature selection scheme to filter out WAPs not providing valuable
features for localization. First, we can filter out WAPs undetectable on a
target floor in a large-scale multibuilding and multifloor fingerprint database
or WAPs no longer active on a target time slot in a dynamic fingerprint database
providing multiple time slices over a long period of time, which are superfluous
for the learning process. Then, we can additionally filter out WAPs based on
their activities and similarity level as described in
Algorithm~\ref{alg:column-selection}. Applying
Algorithm~\ref{alg:column-selection} to the feature matrix
$\vect{X}{\in}\mathbb{R}^{N{\times}W}$, we can reduce its number of columns from
$W$ to $V$, where the similarity threshold value of 3 in its definition and the
comparison threshold value of 0.85 are determined based on the experiments with
the two databases discussed in Section~\ref{sec:exp-results}.

\begin{figure}[!htbp]
  \begin{center}
    \subfigure[]{\includegraphics[width=.9\linewidth,trim=15 18 20 30,clip]{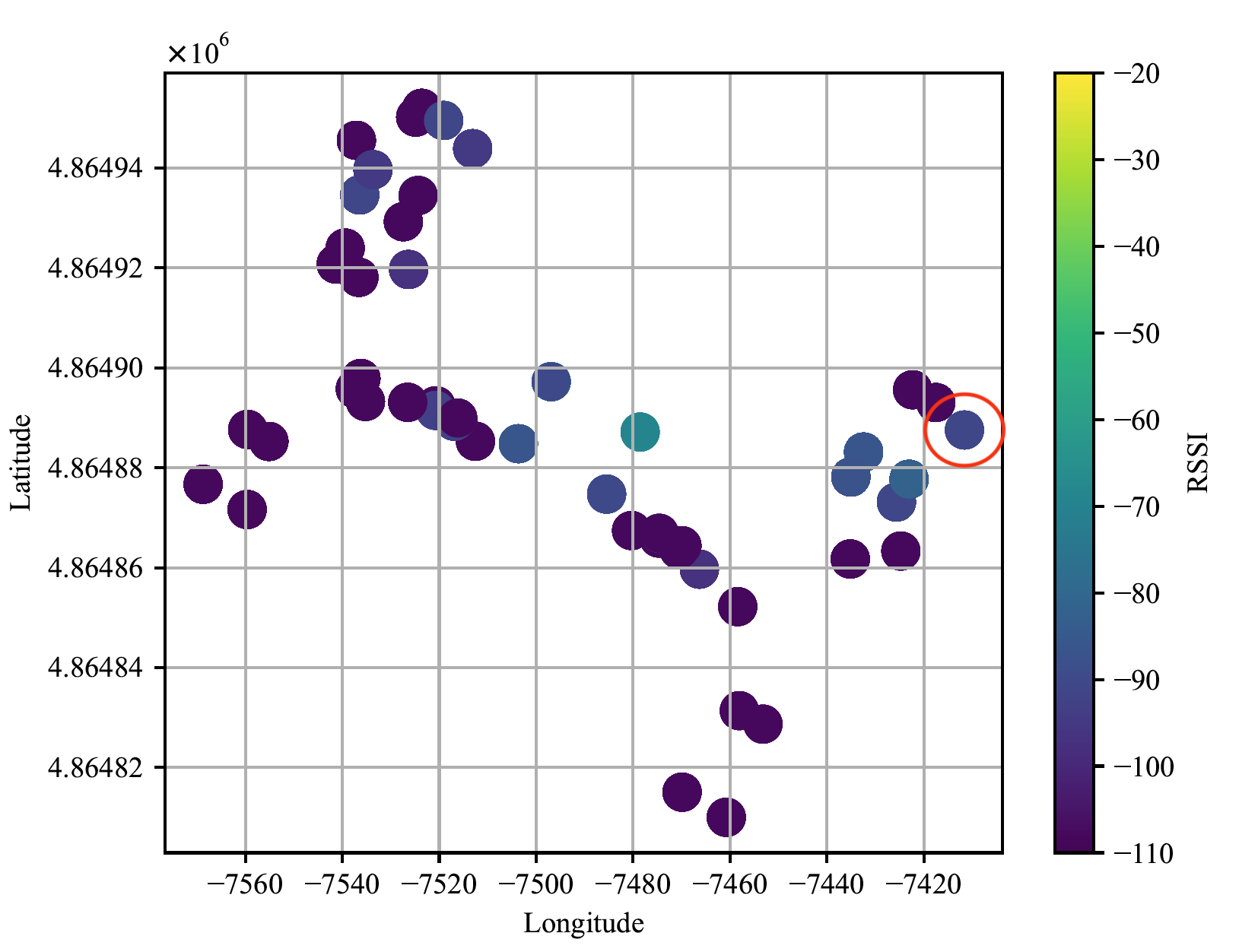}}
    \vspace{-5pt}%
    \subfigure[]{\includegraphics[width=.9\linewidth,trim=15 16 20 30,clip]{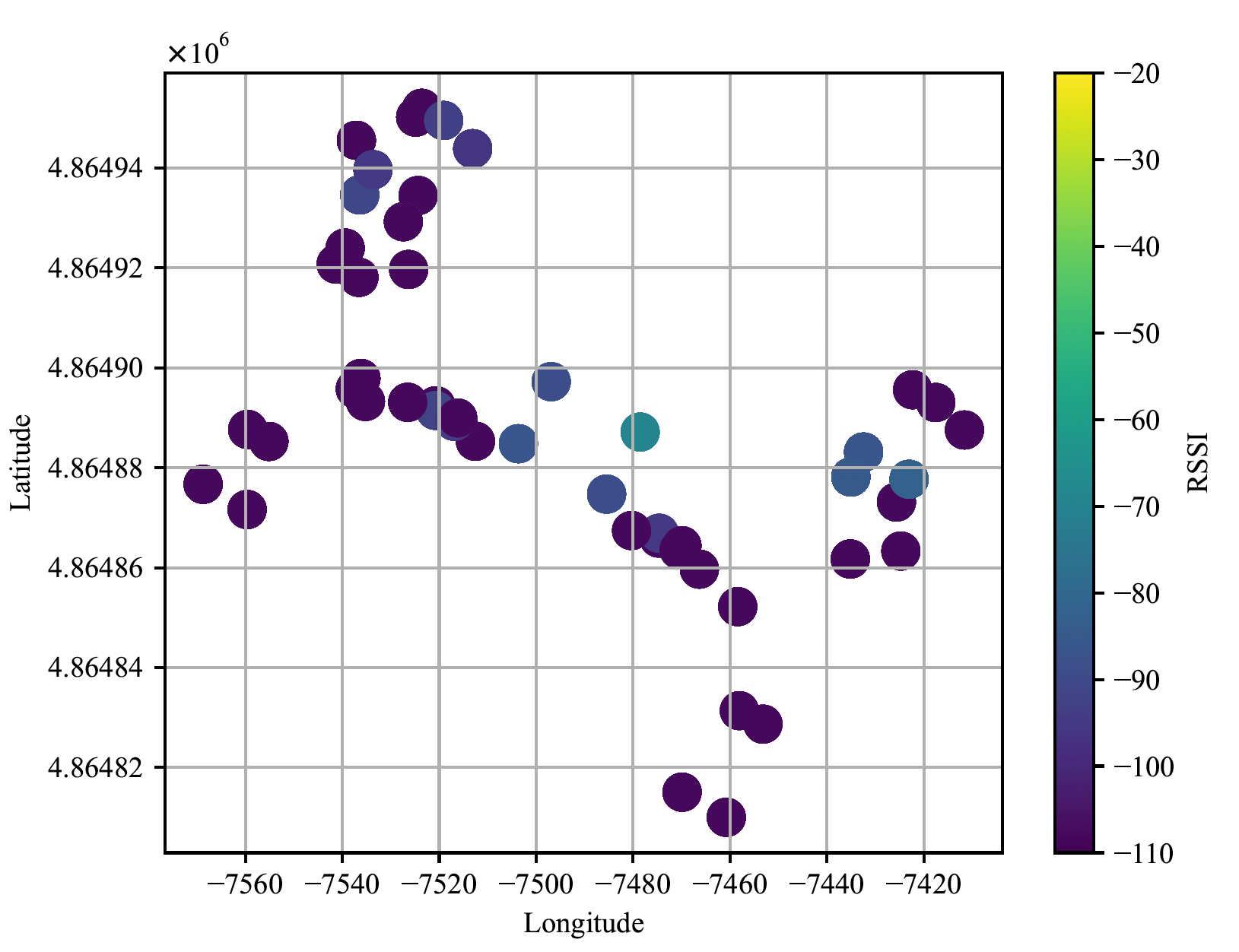}}
  \end{center}
  \caption{Comparison of RSSIs from (a) WAP222 and (b) WAP223, where the red
    circle highlights the difference between them.}
  \label{fig:WAP222-WAP223}
\end{figure}

Note that, because the WAP-based feature selection given in
Algorithm~\ref{alg:column-selection} targets resource-constrained IoT devices,
we intentionally limit its complexity during its design.
Fig.~\ref{fig:WAP222-WAP223} illustrates Algorithm~\ref{alg:column-selection}
with the RSSIs measured on the Floor 3 of Building 1 of the UJIIndoorLoc
database~\cite{UJI} as an example, where the variance of the RSSI vector for
WAP223 is found to be greater than that of WAP222. As described in
Algorithm~\ref{alg:column-selection}, we can construct the difference vector
$\vect{\Delta}$ as a vector of element-wise absolute values of the difference
between the two RSSI vectors. As shown in Fig.~\ref{fig:WAP222-WAP223}, only the
RSSI value highlighted by the red circle at the upper right corner is
significantly different, so more than $85\%$ of the elements in $\vect{\Delta}$
are less than 3 (i.e, $\eta{>}0.85$). Next, the index $k$ of the maximum value
in $\vect{\Delta}$ is obtained, and the RSSI values at this index for WAP223 and
WAP222 are compared. As indicated by the brighter color of the RSSI sample in
the red circle of Fig.~\ref{fig:WAP222-WAP223}, the RSSI value of WAP222 is
greater, so the column belonging to WAP223 is filtered out.

%%%
\SetKwFunction{NC}{NumberOfColumns}%
\SetKwFunction{COL}{Column}%
\begin{algorithm}[!htb]
  \caption{WAP-based feature selection.}
  \label{alg:column-selection}
  \footnotesize%
  \KwData{The feature part $\vect{X}{\in}\mathbb{R}^{N{\times}W}$ of a training
    dataset $\mathcal{D}=\left(\vect{X},\vect{y}\right)$, where $X_{i,j}$ is the
    RSSI from $j$th WAP measured at $i$th RP; a target number of columns
    $V({<}W)$.}%
  \KwResult{A feature part with $V$ columns.}%

  Sort the columns of $\vect{X}$ based on column-wise variances in a decreasing
  order\; $c \leftarrow W$\;
  
  \For{$j=1$ \KwTo $W{-}1$}{ $\vect{\Delta} \leftarrow
    \big|\COL{$\vect{X},j$} - \COL{$\vect{X},j{+}1$}\big|$\tcc*[r]{element-wise
      absolute values}
    $\eta \leftarrow \slfrac{\left(\text{\# of elements of $\vect{\Delta}$ less
          than or equal to 3}\right)}{N}$\; \uIf{$\eta = 1$}{ Remove $(j{+}1)$th
      column from $\vect{X}$\; $c \leftarrow c - 1$\; }
    \ElseIf{$\eta \geq 0.85$}{ $k \leftarrow$ row index of the maximum element
      of $\vect{\Delta}$\; \eIf{$X_{k,j} < X_{k,j+1}$}{ Remove $j$th column from
        $\vect{X}$\; }{ Remove $(j{+}1)$th column from $\vect{X}$\; }
      $c \leftarrow c - 1$\; } \If{$c{=}V$}{ \textbf{break}\; } }
  % $\mathcal{\tilde{D}'} = \{(\vect{x}_{i,j},\vect{y}_{i,j})\}\in \mathbb{R}^{N\times W}$;
\end{algorithm}
%%%
%%%
%%%

\subsubsection{RP-Based Inducing Point Selection}
\label{sec:rp-filtering}
Various SGP algorithms have been proposed for the selection of inducing points,
which can be chosen from the original training data or optimized as part of the
hyperparameters~\cite{quinonero2005unifying}. For the latter, typically
iterative algorithms (e.g, Limited-storage Broyden-Fletcher-Goldfarb-Shanno
(L-BFGS) algorithm~\cite{lbfgs}) are used to maximize an objective function
(e.g., marginal likelihood~\cite{SGP_pseudoinput}), which is not suitable for
resource-constrained IoT devices due to the significant computational
overhead. Like the WAP-based feature selection, we also propose a simple,
heuristic RP-based inducing point selection scheme.

Wi-Fi fingerprint databases can be constructed based on diverse RSSI collection
methods, i.e., insourcing, crowdsourcing, and their hybrid.  In the case of
insourcing, participants of a project collect RSSIs at pre-arranged,
regularly-spaced (e.g., grid) RPs repeatedly; on the other hand, in the case of
crowdsourcing, a large number of volunteers collect RSSIs while moving freely
without pre-arranged RPs, resulting in uneven RSSI spatial distributions. As
there are multiple RSSIs measured at the same RP (i.e., insourcing) or the RPs
located in a small area (i.e., crowdsourcing) whose characteristics are similar
to one another, we select only a few of them as inducing inputs.

For computationally-efficient implementation on resource-constrained IoT
devices, we divide the covered area into a small rectangular grid and randomly
select a certain number of inputs from each grid. The details of the selection
of inducing points based on RPs are described in
Algorithm~\ref{alg:ip-selection}, where the number of grid cells $L$ and the
threshold $\eta$ are design parameters. The algorithm gives the number of
inducing points $M{=}|\mathcal{Z}|$, which is less than or equal to
$L{\times}\eta$.
%%%
\begin{algorithm}[!htb]
  \caption{RP-based inducing point selection.}
  \label{alg:ip-selection}
  \footnotesize%
  \KwData{A training dataset
    $\mathcal{\tilde{D}}{=}\{(\vect{x}_{i},\vect{y}_{i})\}^{N}_{i=1}$ where
    $\vect{y}_{i}$ is the 2D coordinates of the RP for an RSSI vector
    $\vect{x}_{i}$; a set of rectangular grid cells
    $\mathcal{G}{=}\{g_{i}\}^{L}_{i=1}$; a threshold $\eta$ for the maximum
    number of inputs per grid cell.}%
  \KwResult{A set of inducing points $\mathcal{Z}$.}%
  Initialize subsets $\mathcal{Z}_{l}~(l{=}1,{\ldots},L)$
  s.t. $x_{i}{\in}\mathcal{Z}_{l}$ if $\vect{y}_{i}{\in}g_{l}$\;%
  \For{$l=1$ to $L$}{%
    \If{$|\mathcal{Z}_{l}|{>}\eta$}{Remove $|\mathcal{Z}_{l}|{-}\eta$ elements
      from $\mathcal{Z}_{l}$ randomly\;}%
  }%
  $\mathcal{Z}{=}\bigcup_{l=1}^{L}\mathcal{Z}_{l}$\;
\end{algorithm}
%%%

\subsection{Decentralized Indoor Localization Framework}
\label{sec:rt-iot-localization}
\begin{figure}[tb]
  \begin{center}
    \includegraphics[width=0.3\linewidth,angle=270]{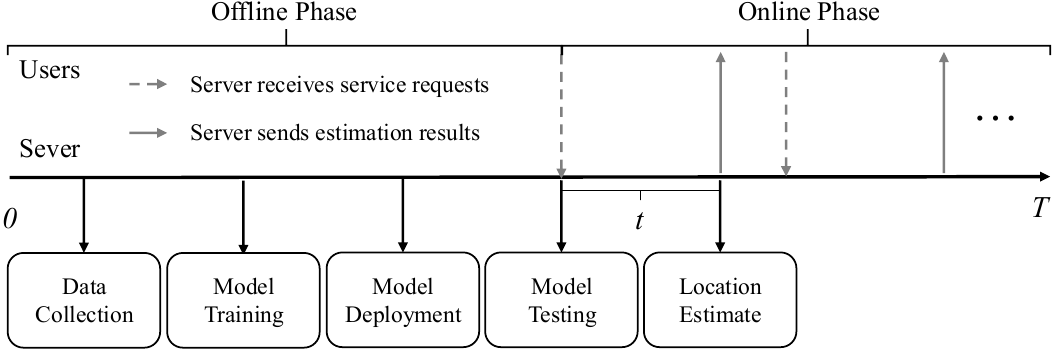}\\
    \vspace{5pt}%
    {\scriptsize (a)}\\
    \vspace{5pt}%
    \includegraphics[width=.9\linewidth]{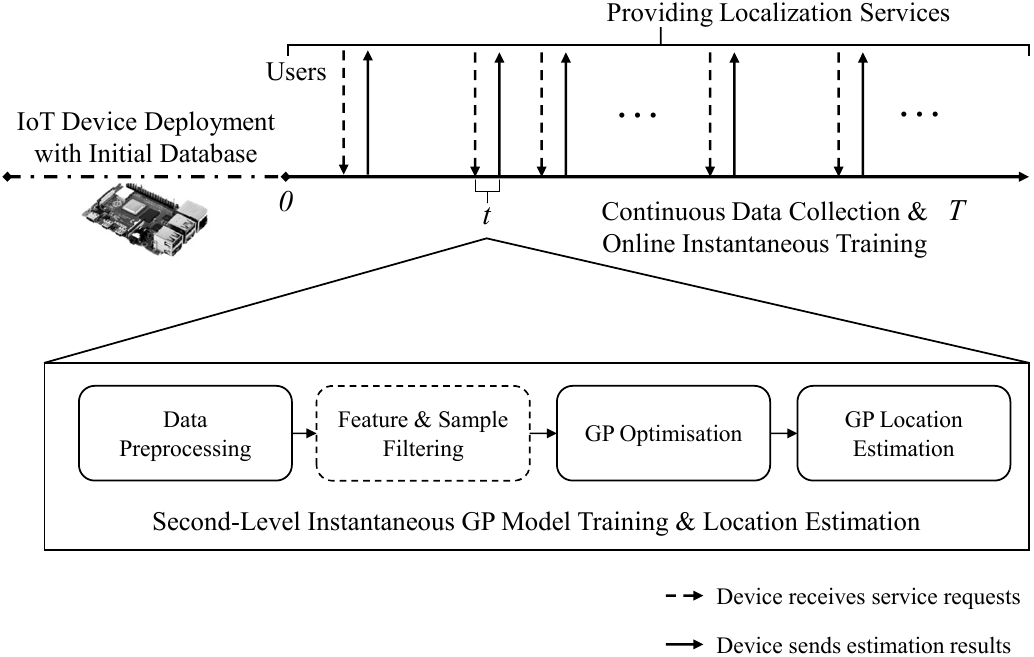}\\
    {\scriptsize (b)}
  \end{center}
  \caption{Comparison of (a) the conventional centralized and (b) the proposed
    decentralized indoor localization frameworks.}
  \label{fig:localization-frameworks}
\end{figure}

%%%
Fig.~\ref{fig:localization-frameworks} shows two different frameworks of indoor
localization, i.e., the conventional, two-phase framework based on a centralized
server and the newly-proposed, decentralized framework based on models deployed
on IoT devices. As shown in Fig.~\ref{fig:localization-frameworks}~(a), the
conventional framework relies on a centralized server and is based on two
separate phases of operation, which makes it difficult to adapt to the
time-varying nature of fingerprint statistics through frequent retraining of the
deployed model. In the newly-proposed framework shown in
Fig.~\ref{fig:localization-frameworks}~(b), the two separate phases of operation
are interleaved and tightly integrated into a unified workflow to provide
continuous data collection and online instantaneous training thanks to the
real-time-trainability of the proposed SGP-RI model, which has much lower
computational complexity than a model based on standard GP with original inputs.

For the proposed indoor localization framework, the floor-level database can be
initiated with a limited number of samples from an existing database covering
the whole multifloor buildings under service or constructed with newly-measured
samples on-site at deployed IoT devices. Once the indoor localization service is
activated, the database is continually expanded and updated through
crowdsourcing or by integrating unlabeled samples from the users of the location
service (e.g., based on semi-supervised learning~\cite{li23:exp-rssi}).

Note that the advantages of the proposed indoor localization framework over the
conventional one are two-fold: First, it can significantly reduce time as well
as labor cost for the construction of a database during the online phase of the
conventional framework and thereby accelerate the deployment and operation of
indoor localization service. Second, through the unified workflow integrating
the separate offline and online phases, both model and database under the
proposed framework are updated continuously unlike those under the conventional
one, which enables location estimation to better reflect the time-varying nature
of fingerprint statistics.
%%%
%%%

\section{Experimental Results}
\label{sec:exp-results}
%%%
To evaluate the performance and computational/data efficiency of the SGP-RI
model, we carried out a series of experiments for both single-building,
single-floor indoor localization with the Xi'an Jiaotong-Liverpool University
(XJTLU) dynamic database~\cite{tang2024static} and multibuilding, multifloor
indoor localization with the UJIIndoorLoc database~\cite{UJI}.

For comparison with reference models under the conventional, two-phase indoor
localization framework based on a centralized server, we run models on two
distinct platforms: (1) a server equipped with an AMD Ryzen 7 5800X processor,
an RTX~3060 Ti GPU, \SI{16}{\giga\byte} of RAM, and \SI{1}{\tera\byte} of
storage, and (2) a Raspberry Pi~4B---as an IoT device---featuring a Cortex-A72
CPU, \SI{4}{\giga\byte} of RAM, and \SI{16}{\giga\byte} of storage.

As for the performance metrics, we adopted the following:
%%%
\begin{itemize}[leftmargin=*]
\item \textit{2D error} for single-floor indoor localization.
\item \textit{3D error} for multibuilding, multifloor indoor localization.
\item \textit{Training time} of a model on a given platform as a measure of
  computational efficiency.
\item \textit{Model sparsity}, the percentage ratio of the number of inducing
  points to the size of the training dataset (i.e., $100{\times}\frac{M}{N}$) in
  approximating GP by SGP-RI~\cite{SGP_pseudoinput}.
\end{itemize}
%%%
Note that the 2D and 3D Errors are related as follows:
%%%
\begin{equation}
  \text{3D Error} = (1 - h_{B}) \times p_{B} + (1 - h_{F}) \times p_{F} + \text{2D Error},
  \label{eq:3d-error}
\end{equation}
%%%
where $h_{B}$ and $h_{F}$ are building and floor hit rates, and $p_{B}$ and
$p_{F}$ are penalties for the errors in building and floor identification that
are set to \SI{50}{\m} and \SI{4}{\m}, respectively~\cite{UJI}.

\subsection{Single-Building and Single-Floor Indoor Localization}
\label{sec:sbsf-localization}
\begin{figure}[tb]
  \begin{center}
    \includegraphics[width=\linewidth]{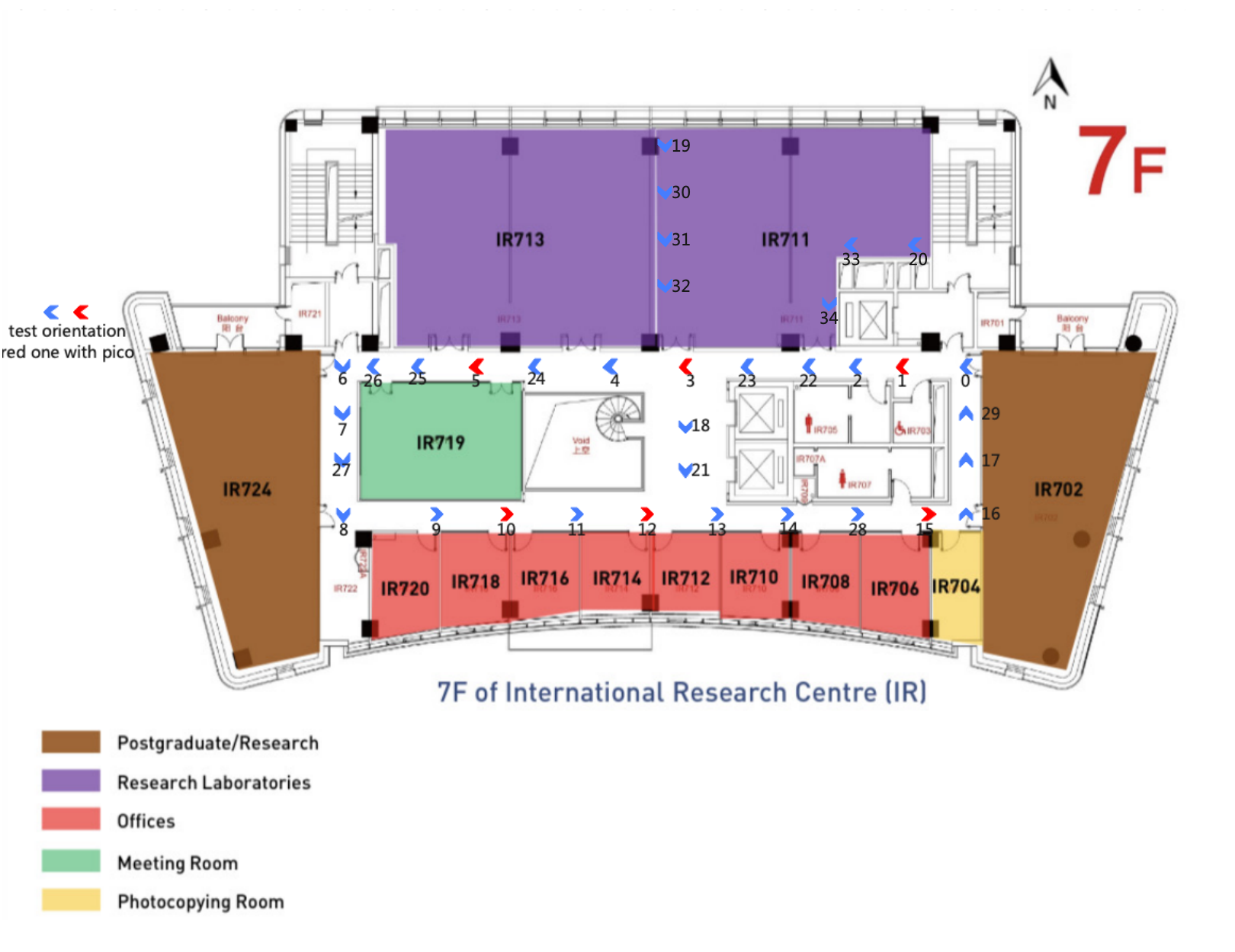}%
    \vspace{-5pt}%
  \end{center}
  \caption{RP distribution on the 7th floor of the XJTLU International Research
    Centre, where the RPs with Raspberry Pi Pico Ws are marked in
    red~\cite{tang2024static}.}
  \label{fig:IR-map}
\end{figure}

%%%
We select the XJTLU dynamic database~\cite{tang2024static} for the performance
evaluation of localization models in single-building and single-floor indoor
localization under dynamic as well as static scenarios, which covers three
floors of the International Research Centre located on XJTLU South Campus, whose
total area is \SI{1200}{\meter\squared}. The Wi-Fi 2.4/5-\si{\GHz} RSSI
fingerprints in the database were measured at 101 RPs---i.e., 28, 35, and 38 RPs
on the sixth, seventh, and eighth floors, respectively---spaced about \SI{3}{\m}
from each other over 44 days by not only tens of surveyors using smartphones and
laptops for daily measurements but also Raspberry Pi Pico Ws equipped with
2.4-\si{\GHz} Wi-Fi interface deployed on the walls for hourly measurements as
shown in Fig.~\ref{fig:IR-map}; 466 WAPs had been detected during the whole
measurement period. For the experiments, we split the XJTLU dynamic database
into a training dataset based on the data measured during the first 24 days and
a test dataset based on the data measured during the remaining 20 days.
%%%
%%%

As for the proposed SGP-RI model, we use the Rational Quadratic (RatQuad)
kernel~\cite{GPML} for all the experiments:
%%%
\begin{equation}
  \label{eq:ratquad-kernel}
  k(\vect{x}_i, \vect{x}_j) = \left(1 + \frac{\|\vect{x}_i - \vect{x}_j\|^2}{2\alpha l^2}\right)^{-\alpha},
\end{equation}
%%%
where $\alpha$ is the scale mixture parameter and $l$ is the
length-scale~\cite{GPML}, whose values are set to 2 and 10, respectively. Note
that we can use different kernels with different parameter values for target
floors to optimize the localization performance in actual deployment because the
performance of GP/SGP-RI models highly depends on the choice of kernels and
their parameter values.

For comparative analyses, we also consider reference models during the
experiments. For NN-based models, we choose simplified CNN and DNN models
tailored for classification. The structure and parameters of DNN and CNN
received inspiration from~\cite{DNN} and~\cite{CNNLoc}.  Both use Stacked Auto
Encoder (SAE) and use, as activation functions, Exponential Linear Unit (ELU)
and Rectified Linear Unit (ReLU), respectively. The DNN includes 4 layers for
classification (CLS), the CNN uses Convolution for 1D (Conv1D), and the network
architectures are summarized in Tables~\ref{tab:dnn_structure}
and~\ref{tab:cnn_structure}; simple adjustments were also made to take into
account database differences. As for conventional algorithms, we consider
$k$-Nearest Neighbors ($k$-NN) and Random Forests (RFs)~\cite{geurts06:_extrem}
for regression, where the number of neighbors for $k$-NN and the number of
decision trees are set to 20 and 100, respectively. In addition, we include the
Microsoft EdgeML Bonsai~\cite{kumar2017resource} as a compact lightweight
baseline for edge-oriented deployment, which is based on a single, shallow
sparse tree.
%%% 
\begin{table}[htbp]
  \caption{DNN network architecture.}
  \label{tab:dnn_structure}
  \begin{center}
    \begin{tabular}{lcc}
      \hline
      Layer & Activation Function & Parameters \\
      \hline
      SAE Encoder & ELU & 465-232-116 \\
      SAE Decoder & ELU & 116-232-465 \\
      CLS Layer 1 & ELU & 512 \\
      CLS Layer 2 & ELU & 512 \\
      CLS Layer 3 & ELU & 512 \\
      CLS Layer 4 & ELU & 512 \\
      Output Layer & - & 35 \\
      \hline
    \end{tabular}
  \end{center}
\end{table}
%%%
%%%
\begin{table}[htbp]
  \caption{CNN network architecture.}
  \label{tab:cnn_structure}
  \begin{center}
    \begin{tabular}{lcc}
      \hline
      Layer & Description & Parameters \\
      \hline
      SAE Encoder & ReLU & 465-232-116 \\
      SAE Decoder & ReLU & 116-232-465 \\
      Conv1D Layer 1 & ReLU & kernel size 22, output channel 99 \\
      % Activation & &ReLU \\
      Conv1D Layer 2 & ReLU & kernel size 22, output channel 66 \\
      % Activation & &ReLU  \\
      Flatten & -  & - \\
      Output Layer & - & 35 \\
      \hline
    \end{tabular}
  \end{center}
\end{table}
%%%

\subsubsection{Experiments Based on The Sever}
\label{sec:exper-based-sever}
%%% 
Before evaluating the realistic performance of the models on the
resource-constrained IoT device for the proposed, decentralized indoor
localization framework, we first assess their absolute performance on the GPU
server with plenty of computational resources for the conventional, two-phase and
centralized indoor localization framework.

From the results summarized in Table~\ref{tab:result-sever}, we observe that the
proposed SGP-RI model can strike the right balance between localization
performance (i.e., 2D Error) and computational efficiency (i.e., Training Time),
especially compared to the GP model. In the case of the SGP-RI model with the
model sparsity of 50\%, for example, we can reduce the training time of the GP
model by more than 50\% at the slight increase of 2D error by about 9\%, while
the SGP-RI model with the model sparsity of 30\% further reduces the training
time to \SI{5.00}{\s} with a still competitive 2D error of \SI{6.44}{\m}. The
Bonsai baseline reaches a 2D error of \SI{7.28}{\m} and requires
\SI{22.74}{\s} for training at the model sparsity of 80\%, which indicates that
its compact model structure does not translate into a better accuracy-latency
trade-off in this experiment. Note that, though computationally highly
efficient, RF and $k$-NN algorithms cannot provide decent localization
performance unlike the SGP-RI model.

\begin{figure}[tb]
  \begin{center}
    \includegraphics[width=.8\linewidth,clip]{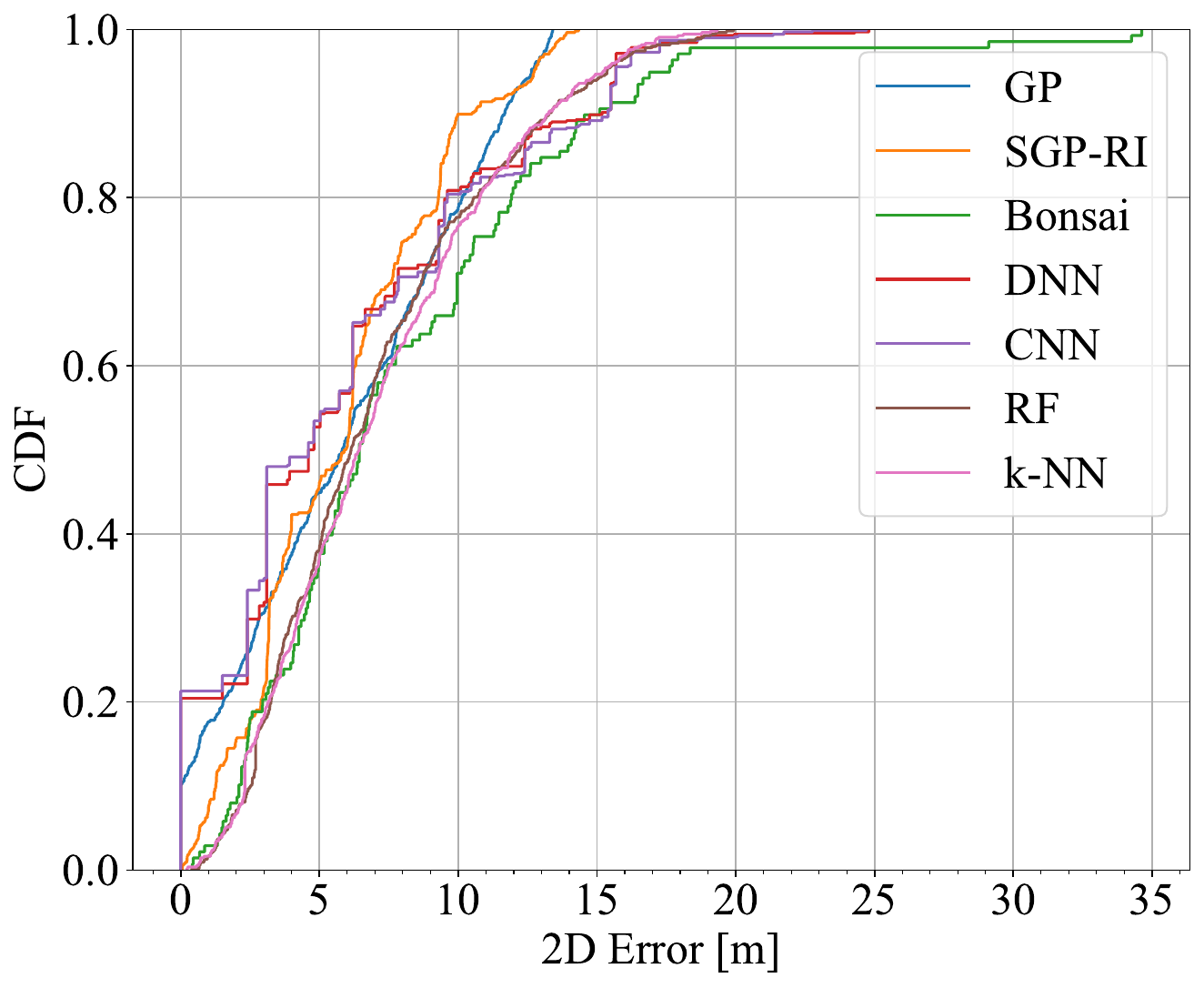}
  \end{center}
  \caption{CDF of the 2D error of the experiment based on sever.}
  \label{fig:cdf}
\end{figure}

To gain a deeper insight into the error distribution, we generated both
Cumulative Distribution Function (CDF) curves for the errors as shown in
Fig.~\ref{fig:cdf}. Assuming a maximum
error tolerance of \SI{5}{\m}, it was observed that the probability of staying
within this tolerance is only achieved when using inducing points larger than
$25\%$. Also, the maximum value of the proposed model is less than \SI{15}{\m}, 
while the other models have a maximum value of less than \SI{25}{\m}.
%%%
\begin{table}[]
  \centering
  \begin{threeparttable}
    \caption{Experimental results based on the server.}
    \label{tab:result-sever}
    \begin{tabular}{lccc}
      \hline
      Model     & 2D Error {[}m{]} & Training Time {[}s{]} & Model Sparsity \\
      \hline
      GP        & \textbf{5.32} & 12.79             & --- \\
      SGP-RI    & 5.80          & 6.08              & 50\% \\
      SGP-RI    & 5.96          & 5.52              & 40\% \\
      SGP-RI    & 6.44          & 5.00              & 30\% \\
      DNN       & 5.86          & 17.18\tnote{*}    & --- \\
      CNN       & 5.87          & 12.06\tnote{*}    & --- \\
      RF        & 7.00          & 1.11              & --- \\
      $k$-NN    & 7.12          & \textbf{0.07}     & --- \\
      Bonsai    & 7.28          & 22.74             & 80\% \\
      \hline
    \end{tabular}%
    \begin{tablenotes}
    \item[*] GPU enabled.
    \end{tablenotes}
  \end{threeparttable}
\end{table}
%%%
%%%
%%%

\subsubsection{Experiments Based on The IoT System Device}
\label{sec:exper-based-iot}
%%% 
To complete the experiments on IoT system devices, we used a Raspberry Pi 4B to
simulate WAPs with redundant arithmetic in the IoT system.  Due to the lack of
powerful GPU on the Raspberry Pi 4B, we could not train and run the DNN and CNN
models.  Given the Raspberry Pi's lack of active cooling and the risk of
overheating during prolonged or consecutive testing, all the experiments were
conducted in an environment maintained at approximately
\SI{25}{\degreeCelsius}. Adequate time was allowed for the Raspberry Pi to cool
to room temperature before subsequent testing commenced. Please note that
fluctuations in training time are strongly influenced by temperature, so orders
of magnitude of training time are meaningful, not exact times. In addition, if
GP is deployed, a Raspberry Pi 4B with at least \SI{64}{\giga\byte} of storage,
and active cooling to ensure that operating temperatures are always below
\SI{50}{\degreeCelsius} is required, a condition that is inconsistent with other
models mentioned in Section~\ref{sec:exp-results}.

The experimental results based on the Raspberry Pi 4B are summarized in
Table~\ref{tab:result-IoT}. Despite the low computational capability on the
device, the proposed SGP-RI model successfully completed training within
\SI{30}{\s} and delivered reliable accuracy. This demonstrates the algorithm's
efficacy in resource-constrained environments, such as IoT devices, where it can
maintain efficiency and precision. The Bonsai baseline can also be trained on
the Raspberry Pi 4B, but it reaches a 2D error of \SI{7.34}{\m} and requires
\SI{120.47}{\s} for training at the model sparsity of 80\%. Compared with all
three SGP-RI settings, Bonsai has both higher localization error and longer
training time, suggesting that the proposed SGP-RI model is more suitable for
frequent local retraining under the proposed decentralized framework.
%%%
\begin{table}[]
  \centering
  \begin{threeparttable}
    \caption{Experimental results based on the IoT system device.}
    \label{tab:result-IoT}
    \begin{tabular}{lccc}
      \hline
      Model     & 2D Error {[}m{]} & Training Time {[}s{]} & Model Sparsity \\ \hline
      GP\tnote{*}     & \textbf{5.44} & 96.34        & ---  \\ 
      SGP-RI & 5.84          & 24.03        & 50\%  \\
      SGP-RI & 5.96                   & 20.45        & 40\%  \\
      SGP-RI & 6.50                   & 18.20        & 30\%  \\ 
      RF     & 7.08                   & 2.62         & --- \\
      $k$-NN    & 7.10                   & \textbf{1.17}         & --- \\
      Bonsai & 7.34                   & 120.47       & 80\%  \\
      \hline
    \end{tabular}%
    \begin{tablenotes}
    \item[*] GP's experiment equipment and requirements differ from those of other models as detailed in Section~\ref{sec:exper-based-iot}.
    \end{tablenotes}
  \end{threeparttable}
\end{table}
%%%

\subsubsection{Experiments Under A Dynamic Localization Scenario}
\label{sec:exper-dynam-model}
%%%
As demonstrated by the experimental results in
Section~\ref{sec:exper-based-iot}, the real-time-trainability of the proposed
SGP-RI model on IoT devices provides the continuous data collection and online
instantaneous training under the proposed indoor localization framework shown in
Figure~\ref{fig:localization-frameworks}~(b) and thereby enables the model to
adapt to the dynamic and fluctuating indoor electromagnetic environment, which
is not possible under the conventional, two-phase indoor localization framework
based on a centralized server due to the prohibitive cost of retraining.

To demonstrate the advantage of the proposed indoor localization model and
framework over the conventional ones, we simulated the following post-deployment
scenario: For the DNN and CNN models, the training dataset is still based on the
measurements during the first 24 days, but the test dataset is now divided into
four groups of 5-day measurements each, for each of which we calculate the 2D
error separately. For the SGP-RI, RF, and $k$-NN models, the initial training
dataset is based on the measurements during the first 24 days, but the four
groups of the test dataset are sequentially moved from the test dataset to the
training dataset, which simulates those models operating under the proposed
indoor localization framework based on frequent retraining with new sets of data
over time. The results in Table~\ref{tab:result-grouped} show that the proposed
SGP-RI model provides the best 2D errors over the whole period thanks to its
frequent retraining, highlighting its capability to maintain higher localization
performance in spite of the environmental changes. It is this adaptability that
differentiates the proposed indoor localization model and framework over the
conventional ones.
%%%
\begin{table}[]
  \centering
  \caption{2D Errors under a dynamic localization scenario.}
  \label{tab:result-grouped}
  \begin{tabular}{lcccc}
    \hline
    \multirow{2}{*}{Model} & \multicolumn{4}{c}{2D Error for Each Test Period {[}m{]}}                                                                               \\ \cline{2-5} 
                           & \multicolumn{1}{l}{1-5} & \multicolumn{1}{l}{6-10} & \multicolumn{1}{l}{11-15} & \multicolumn{1}{l}{16-20} \\ \hline
    DNN                    & 5.58                    & 5.77                     & 5.96                      & 6.12                      \\
    CNN                    & 5.72                    & 5.82                     & 6.05                      & 5.89                      \\
    \textbf{SGP-RI}     & \textbf{5.46}           & \textbf{5.42}            & \textbf{5.64}             &\textbf{5.80}              \\ 
    RF                     & 6.76                    & 6.66                     & 6.81                      & 6.82                      \\ 
    $k$-NN                    & 6.88                    & 6.68                     & 6.90                      & 7.07                      \\ 
    \hline
  \end{tabular}%
\end{table}
%%%

\subsection{Multibuilding and Multifloor Indoor Localization}
\label{sec:mbmf-localization}
%%%
Table~\ref{tab:result-UJI} summarizes the performance of multibuilding and
multifloor indoor localization of the proposed SGP-RI model with the model
sparsity of 50\% (i.e., ``SGP-RI 50\%'') as well as that of the reference models
in the literature based on the UJIIndoorLoc database~\cite{UJI}, which indicates
that the proposed SGP-RI model can also provide indoor localization performance
comparable to that of the state-of-the-art models under a multibuilding and
multifloor environment.

Note that the 3D errors listed in Table~\ref{tab:result-UJI} should be
interpreted as relative indicators of the models' performance because they are
not calculated under the same condition as frequently discussed in the
literature (e.g.,~\cite{DNN,MOGP-sensors}): The top four models from the 2015
EvAAL/IPIN competition~\cite{EvAAL} are evaluated based on the training,
validation, and test datasets of the UJIIndoorLoc database, the last of which,
however, is not publicly available. Therefore, the rest of the models are
evaluated based only on the training and validation datasets of the UJIIndoorLoc
database.\footnote{Most researchers split the validation dataset into new
  validation and test datasets, while, in~\cite{10496456}, the training and
  validation datasets are merged into one common dataset before being split into
  new training, validation, and test datasets.}

Also, the calculation of the 3D error under the new decentralized indoor
localization framework based on the SGP-RI models deployed on IoT devices is
different from that under the conventional, two-phase indoor localization
framework based on a model running on a centralized server. Unlike the
conventional framework, the proposed framework uses a single IoT device for
location service covering a whole or part of a floor, whose location is known
during its deployment. Under this decentralized framework, it is assumed that
the estimation of the building and floor is based on the RSSIs from IoT devices
measured at a user's device, where the strongest-RSSI IoT device can provide the
building and floor information of a user. Considering a large attenuation of
Wi-Fi signals from different buildings, we set $h_B$, the building hit rate, to
1, which is more or less consistent with the building estimation performance of
the state-of-the-art centralized models. As for $h_F$, the floor hit rate, it is
reasonable to assume that an IoT devices with the strongest-RSSI is located at
the same floor as the user or its neighboring floors. Determining $h_{F}$ in a
multibuilding, multifloor environment, therefore, can be transformed into a
binary (i.e., top and bottom floors) or ternary (i.e., all other floors)
classification task. We designed a simple $k$-NN with $k{=}7$ to accomplish the
classification task and evaluated the value of $h_{F}$ using the UJIIndoorLoc
database, whose average is 80\%. More advanced and complex classification
algorithms likely result in higher $h_{F}$ values, of course, but it is not
feasible for resource-constrained IoT devices.
%%%
\begin{table}[]
  \centering
  \caption{3D errors based on the UJIIndoorLoc database.}
  \label{tab:result-UJI}
  \begin{tabular}{clc}
    \hline
    & \multicolumn{1}{c}{Model} & 3D Error {[}m{]} \\ \hline
    \multirow{4}{*}{\begin{tabular}[c]{@{}c@{}}EvAAL\\ IPIN 2015~\cite{EvAAL}\end{tabular}} 
    & RTLS@UM                   & 6.20                       \\
    & ICSL                      & 7.67                        \\
    & HFTS                      & 9.49                        \\
    & MOSAIC                    & 11.64                      \\ \hline
    & \textbf{SGP-RI 50\%}         & \textbf{6.87}                         \\
    & CDAELoc~\cite{10496456}                   & 7.37                               \\
    & SALLoc~\cite{10416861}                    & 8.28                             \\
    & EA-CNN~\cite{alitaleshi2023ea}                    & 8.34                       \\
    & RNN-MOGP Aug~\cite{MOGP-sensors}              & 8.42                         \\
    & RNN~\cite{2021hierarchical}                       & 8.62                         \\
    & CHISEL~\cite{wang2021chisel}                    & 8.80                         \\
    & DNN-DLB~\cite{s21062000}                   & 9.07                          \\
    & Scalable DNN~\cite{DNN}              & 9.29                          \\
    & CNNLoc~\cite{CNNLoc}                    & 11.78                     \\
    & CCpos~\cite{qin2021ccpos}                     & 12.4                              \\ \hline
  \end{tabular}%
\end{table}
%%%

\subsection{Effects of the SGP-RI Threshold Values}
\label{sec:sgp-ri-thresholds}
%%%
% \todos{Tim}{To present the results of the analysis of parameter sensitivity and
%   related discussions to address the comment~\#2 from reviewer~\#1.}%
%%%
To examine the sensitivity of the empirical threshold in
Algorithm~\ref{alg:column-selection}, we repeated the server-side SGP-RI
experiment with the model sparsity fixed at 30\%, varying only the WAP similarity
comparison threshold. As shown in Table~\ref{tab:threshold-sensitivity}, lowering
the threshold to 0.75 reduces the training time but increases the 2D error, while
raising it to 0.95 improves accuracy at the cost of a much longer training time.
The threshold value of 0.85 is therefore retained as a balanced default between
localization accuracy and training efficiency.
%%%
\begin{table}[tb]
  \centering
  \caption{Sensitivity of SGP-RI to the WAP similarity comparison threshold on
    the server.}
  \label{tab:threshold-sensitivity}
  \begin{tabular}{ccc}
    \hline
    Threshold & 2D Error {[}m{]} & Training Time {[}s{]} \\ \hline
    0.75      & 6.58             & 4.61                  \\
    0.85      & 6.44             & 5.00                  \\
    0.95      & \textbf{6.11}    & 9.23                  \\ \hline
  \end{tabular}
\end{table}
%%%

\section{Case Studies}
\label{sec:case-studies}
%%%
The proposed decentralized indoor localization is particularly suitable for the
following deployment scenarios: In the case of \textit{healthcare at
  hospitals/clinics}, the major advantages are improved privacy (especially for
patients) and service continuity because keeping fingerprint databases and
localization models on local IoT nodes helps preserve sensitive trajectory data,
reduces dependence on a central server, and maintains service continuity when
network backhaul is limited or unreliable. The improved privacy is also
important to \textit{multitenant smart environments}, where the sensitive
trajectory data of each tenant could be physically separated into a dedicated
IoT node that can also handle tenant turn over independently of other nodes.

The cases of \textit{dynamic industrial warehouses}, \textit{disaster recovery},
and \textit{underground mining}, on the other hand, are benefited by the
adaptability and reliability of the proposed decentralized indoor localization
that can quickly adapt to layout changes, moving equipment, signal blockage, or
emergency deployments without rebuilding a full centralized database.

\section{Conclusions}
\label{sec:conclusions}
%%%
We propose the decentralized indoor localization based on SGP-RI models deployed
on resource-constrained IoT devices, enabling real-time sensing and training to
better adapt to dynamic and fluctuating indoor electromagnetic environments.
Experimental results from both dynamic and static Wi-Fi fingerprint databases
demonstrate the feasibility of this approach. Notably, the SGP-RI model not only
provides comparable localization performance using a small fraction of the
training data but also greatly reduces training time.

Our work differs from prior work on decentralized indoor localization in that it
provides a comprehensive framework targeting dynamic environments, which covers
all aspects of Wi-Fi fingerprinting from the construction/maintenance of
fingerprint databases to training/retraining of localization models to a unified
workflow integrating the two operational phases of the conventional indoor
localization.

The proposed framework reduces reliance on a centralized server, which could be
a single point of failure, and thereby mitigates service disruptions caused by
malicious attacks. This is achieved through multiple models running on IoT
devices covering only target floors of the whole service area; the same IoT
devices can automatically collect Wi-Fi fingerprints as well, which is the case
for the construction of the XJTLU dynamic database~\cite{tang2024static} and
provides a promising solution for the maintenance and update of fingerprint
databases~\cite{li24:_wi_fi}.

Improving the WAP-based feature selection and the RP-based inducing point
selection could be interesting topics for future work. Given that IoT devices
have been being deployed in the field and integrated into existing Wi-Fi
infrastructure based on centrally-managed WAPs, it is also worth investigating a
hybrid indoor localization framework utilizing both centralized server and
multiple IoT devices running real-time-trainable models to get the benefits of
both centralized and decentralized indoor localization frameworks.

Beyond model-level improvements, the network-wide scalability, coordination
between IoT nodes, and communication overhead of the proposed framework could be
investigated. Large-scale deployments require principled choices of node density
and coverage granularity, handoff and conflict-resolution policies for
overlapping service areas, and lightweight exchange of metadata, model status,
and update summaries rather than raw fingerprints. These system-level issues are
orthogonal to the single-node localization model evaluated here but essential
for campus- or building-scale deployment.
%%%

\balance

%%% References
% % - with BiBTeX
% \bibliographystyle{IEEEtran}%
% \bibliography{IEEEabrv,reference}%

\begin{thebibliography}{10}
\providecommand{\url}[1]{#1}
\csname url@samestyle\endcsname
\providecommand{\newblock}{\relax}
\providecommand{\bibinfo}[2]{#2}
\providecommand{\BIBentrySTDinterwordspacing}{\spaceskip=0pt\relax}
\providecommand{\BIBentryALTinterwordstretchfactor}{4}
\providecommand{\BIBentryALTinterwordspacing}{\spaceskip=\fontdimen2\font plus
\BIBentryALTinterwordstretchfactor\fontdimen3\font minus
  \fontdimen4\font\relax}
\providecommand{\BIBforeignlanguage}[2]{{%
\expandafter\ifx\csname l@#1\endcsname\relax
\typeout{** WARNING: IEEEtran.bst: No hyphenation pattern has been}%
\typeout{** loaded for the language `#1'. Using the pattern for}%
\typeout{** the default language instead.}%
\else
\language=\csname l@#1\endcsname
\fi
#2}}
\providecommand{\BIBdecl}{\relax}
\BIBdecl

\bibitem{s21238086}
T.~Yang, A.~Cabani, and H.~Chafouk, ``A survey of recent indoor localization
  scenarios and methodologies,'' \emph{Sensors}, vol.~21, no.~23, 2021,
  (article number: 8086).

\bibitem{Low-Cost-Method-A-Review}
W.~Liu, Y.~Zhang, Z.~Deng, and H.~Zhou, ``Low-cost indoor wireless fingerprint
  location database construction methods: A review,'' \emph{{IEEE} Access},
  vol.~11, pp. 37\,535--37\,545, 2023.

\bibitem{DNN}
K.~S. Kim, S.~Lee, and K.~Huang, ``A scalable deep neural network architecture
  for multi-building and multi-floor indoor localization based on {Wi-Fi}
  fingerprinting,'' \emph{Big Data Analytics}, vol.~3, no.~1, Apr. 2018.

\bibitem{CNNLoc}
X.~Song, X.~Fan, C.~Xiang, Q.~Ye, L.~Liu, Z.~Wang, X.~He, N.~Yang, and G.~Fang,
  ``A novel convolutional neural network based indoor localization framework
  with {WiFi} fingerprinting,'' \emph{{IEEE} Access}, vol.~7, pp.
  110\,698--110\,709, 2019.

\bibitem{2021hierarchical}
A.~E. Ahmed~Elesawi and K.~S. Kim, ``Hierarchical multi-building and
  multi-floor indoor localization based on recurrent neural networks,'' in
  \emph{Proc. 2021 {CANDARW}}, 2021, pp. 193--196.

\bibitem{Transformers-indoor-localization}
S.~M. Nguyen, D.~V. Le, and P.~J. Havinga, ``Seeing the world from its words:
  All-embracing transformers for fingerprint-based indoor localization,''
  \emph{{Pervasive Mob. Comput.}}, vol. 100, p. 101912, 2024.

\bibitem{10601173}
S.~Wang, S.~Zhang, J.~Ma, and O.~A. Dobre, ``Graph neural network-based {WiFi}
  indoor localization system with access point selection,'' \emph{{IEEE}
  Internet Things J.}, pp. 1--1, 2024.

\bibitem{CNN-Transformer}
N.~Savin, ``Exploring indoor localization with transformer-based models: A
  {CNN}-transformer hybrid approach for {WiFi} fingerprinting,'' in \emph{Proc.
  39th {Twente} Student Conference on {IT}}.\hskip 1em plus 0.5em minus
  0.4em\relax Twente, Netherlands: University of Twente, July 2023, pp. 1--5.

\bibitem{OpenWrt}
\BIBentryALTinterwordspacing
``{OpenWrt} project.'' [Online]. Available: \url{https://openwrt.org/start}
\BIBentrySTDinterwordspacing

\bibitem{venkatesh2018iot}
K.~Venkatesh, P.~Rajkumar, S.~Hemaswathi, and B.~Rajalingam, ``{IoT} based home
  automation using {Raspberry Pi},'' \emph{J. Adv. Res. Dyn. Control Syst},
  vol.~10, no.~7, pp. 1721--1728, 2018.

\bibitem{8454750}
Z.~Kasmi, N.~Guerchali, A.~Norrdine, and J.~H. Schiller, ``Algorithms and
  position optimization for a decentralized localization platform based on
  resource-constrained devices,'' \emph{{IEEE} Trans. Mobile Comput.}, vol.~18,
  no.~8, pp. 1731--1744, 2019.

\bibitem{kim25:_dual_prong_solut_accur_decen_tag_system}
M.~Kim, S.~Bahrami, and W.~Hong, ``A dual-pronged solution for accurate
  decentralized tag systems: {RSSD}-based {DoA} and tag design,'' \emph{{IEEE}
  Internet Things J.}, vol.~12, no.~22, pp. 47\,896--47\,911, 2025.

\bibitem{ye22:_edgel}
Q.~Ye, H.~Bie, K.-C. Li, X.~Fan, L.~Gong, X.~He, and G.~Fang, ``{EdgeLoc}: A
  robust and real-time localization system toward heterogeneous {IoT}
  devices,'' \emph{{IEEE} Internet Things J.}, vol.~9, no.~5, pp. 3865--3876,
  2022.

\bibitem{nikitaki12:_decen}
S.~Nikitaki and P.~Tsakalides, ``Decentralized indoor wireless localization
  using compressed sensing of signal-strength fingerprints,'' in \emph{Proc.
  {PM2HW2N '12}}, Paphos, Cyprus, 2012, pp. 37--44.

\bibitem{garg2023sirius}
N.~Garg and N.~Roy, ``Sirius: A self-localization system for
  resource-constrained {IoT} sensors,'' in \emph{Proc. {MobySys} '23}, 2023,
  pp. 289--302.

\bibitem{8953024}
A.~Christidis, R.~Davies, and S.~Moschoyiannis, ``Serving machine learning
  workloads in resource constrained environments: a serverless deployment
  example,'' in \emph{in Proc. 2019 {SOCA}}, 2019, pp. 55--63.

\bibitem{9149180}
B.~Sliwa, N.~Piatkowski, and C.~Wietfeld, ``{LIMITS}: Lightweight machine
  learning for {IoT} systems with resource limitations,'' in \emph{Proc. 2020
  {ICC}}, 2020, pp. 1--7.

\bibitem{8054360}
F.~Zhou, K.~Lin, A.~Ren, D.~Cao, Z.~Zhang, M.~U. Rehman, X.~Yang, and
  A.~Alomainy, ``{RSSI} indoor localization through a {Bayesian} strategy,'' in
  \emph{Proc. 2017 {IAEAC}}, 2017, pp. 1975--1979.

\bibitem{moghtadaiee2014design}
V.~Moghtadaiee and A.~G. Dempster, ``Design protocol and performance analysis
  of indoor fingerprinting positioning systems,'' \emph{Physical
  Communication}, vol.~13, pp. 17--30, 2014.

\bibitem{li24:_wi_fi}
S.~Li, Z.~Tang, K.~S. Kim, and J.~S. Smith, ``On the use and construction of
  {Wi-Fi} fingerprint databases for large-scale multi-building and multi-floor
  indoor localization: A case study of the {UJIIndoorLoc} database,''
  \emph{Sensors}, vol.~24, no.~12, 2024, (article number: 3827).

\bibitem{UJI}
J.~Torres-Sospedra, R.~Montoliu, A.~Martínez-Usó, J.~P. Avariento, T.~J.
  Arnau, M.~Benedito-Bordonau, and J.~Huerta, ``{UJIIndoorLoc}: A new
  multi-building and multi-floor database for {WLAN} fingerprint-based indoor
  localization problems,'' in \emph{Proc. 2014 {IPIN}}, 2014, pp. 261--270.

\bibitem{GPML}
C.~E. Rasmussen and C.~K.~I. Williams, \emph{{Gaussian} Processes for Machine
  Learning}, ser. Adaptive Computation and Machine Learning.\hskip 1em plus
  0.5em minus 0.4em\relax {MIT Press}, 2006.

\bibitem{tran2016variational}
D.~Tran, R.~Ranganath, and D.~M. Blei, ``The variational {Gaussian} process,''
  \emph{arXiv preprint arXiv:1511.06499}, 2015.

\bibitem{MOGP-sensors}
Z.~Tang, S.~Li, K.~S. Kim, and J.~S. Smith, ``Multi-dimensional {Wi-Fi}
  received signal strength indicator data augmentation based on {Multi-Output
  Gaussian Process} for large-scale indoor localization,'' \emph{Sensors},
  vol.~24, no.~3, 2024, (article number: 1026).

\bibitem{quinonero2005unifying}
J.~Quinonero-Candela and C.~E. Rasmussen, ``A unifying view of sparse
  approximate {Gaussian} process regression,'' \emph{{J. Mach. Learn. Res.}},
  vol.~6, pp. 1939--1959, 2005.

\bibitem{SGP_pseudoinput}
E.~Snelson and Z.~Ghahramani, ``Sparse {Gaussian} processes using
  pseudo-inputs,'' in \emph{Advances in Neural Information Processing Systems},
  Y.~Weiss, B.~Sch\"{o}lkopf, and J.~Platt, Eds., vol.~18.\hskip 1em plus 0.5em
  minus 0.4em\relax MIT Press, 2005.

\bibitem{lbfgs}
J.~Nocedal, ``Updating quasi-newton matrices with limited storage,''
  \emph{Mathematics of computation}, vol.~35, no. 151, pp. 773--782, 1980.

\bibitem{li23:exp-rssi}
S.~Li, Z.~Tang, K.~S. Kim, and J.~S. Smith, ``Exploiting unlabeled {RSSI}
  fingerprints in multi-building and multi-floor indoor localization through
  deep semi-supervised learning based on mean teacher,'' in \emph{Proc.
  {CANDAR} 2023}, 2023, pp. 155--160, outstanding paper award.

\bibitem{tang2024static}
Z.~Tang, R.~Gu, S.~Li, K.~S. Kim, and J.~S. Smith, ``Static vs. dynamic
  databases for indoor localization based on {Wi-Fi} fingerprinting: A
  discussion from a data perspective,'' in \emph{Proc. 2024 {ICAIIC}}, 2024,
  pp. 760--765.

\bibitem{geurts06:_extrem}
P.~Geurts, D.~Ernst, and L.~Wehenkel, ``Extremely randomized trees,''
  \emph{Machine Learning}, vol.~63, pp. 3--42, 2006.

\bibitem{kumar2017resource}
A.~Kumar, S.~Goyal, and M.~Varma, ``Resource-efficient machine learning in
  2~{KB} {RAM} for the {Internet of Things},'' in \emph{Proc. {ICML} 2017}, May
  2017, pp. 1--10.

\bibitem{EvAAL}
A.~Moreira, M.~J.~a. Nicolau, F.~Meneses, and A.~Costa, ``{Wi-Fi}
  fingerprinting in the real world - {RTLS\@UM} at the {EvAAL} competition,''
  in \emph{Proc. 2015 {IPIN}}, 2015, pp. 1--10.

\bibitem{10496456}
P.~Zhou, H.~Wang, R.~Gravina, and F.~Sun, ``{WIO-EKF}: Extended {Kalman}
  filtering-based {Wi-Fi} and inertial odometry fusion method for indoor
  localization,'' \emph{{IEEE} Internet Things J.}, pp. 1--1, 2024.

\bibitem{10416861}
S.~L. Ayinla, A.~A. Aziz, and M.~Drieberg, ``{SALLoc}: An accurate target
  localization in {WiFi-Enabled} indoor environments via {SAE-ALSTM},''
  \emph{{IEEE} Access}, vol.~12, pp. 19\,694--19\,710, 2024.

\bibitem{alitaleshi2023ea}
A.~Alitaleshi, H.~Jazayeriy, and J.~Kazemitabar, ``{EA-CNN}: A smart indoor
  {3D} positioning scheme based on {Wi-Fi} fingerprinting and deep learning,''
  \emph{{Eng. Appl. Artif. Intell.}}, vol. 117, p. 105509, 2023.

\bibitem{wang2021chisel}
L.~Wang, S.~Tiku, and S.~Pasricha, ``Chisel: compression-aware high-accuracy
  embedded indoor localization with deep learning,'' \emph{{IEEE} Embedded
  Syst. Lett.}, vol.~14, no.~1, pp. 23--26, 2021.

\bibitem{s21062000}
M.~Laska and J.~Blankenbach, ``Deeplocbox: Reliable fingerprinting-based indoor
  area localization,'' \emph{Sensors}, vol.~21, no.~6, 2021, (article number:
  2000).

\bibitem{qin2021ccpos}
F.~Qin, T.~Zuo, and X.~Wang, ``{CCpos}: {WiFi} fingerprint indoor positioning
  system based on {CDAE-CNN},'' \emph{Sensors}, vol.~21, no.~4, 2021, (article
  number: 1114).

\end{thebibliography}

\end{document}